\documentclass{actasen}

\usepackage{diagbox,multirow,float}
\begin{document}

\setcounter{page}{317}

\Volume{2014}{42}


\runheading{YU Liang-liang et al.}

\title{Investigation of Thermal Inertia and Surface Properties
 for Near-Earth Asteroid (162173) 1999 JU3$^{\dag}~ \!^{\star}$}

\footnotetext{$^{\dag}$ Supported by National Natural Science
Foundation (11273068, 11203087, 10973044, 10933004),
Pilot Project for Layout of Emerging and Cross
Disciplines, Chinese Academy of Sciences (KJZD-EW-Z001), Financial
Item for Astronomy of CAS, Natural Sciences Foundation of Jiangsu
Province (BK2009341), and Asteroid Foundation of Purple Mountain
Observatory.

\hspace*{3mm}Received 2013-04-14

$^{\star}$ A translation of {\it Acta Astron. Sin.~}
Vol. 54, No. 6, pp. 537--548, 2013 \\
\hspace*{5mm}$^{\bigtriangleup}$ yullmoon@pmo.ac.cn\\
\hspace*{5mm}$^{\bigtriangleup}$$^{\bigtriangleup}$ jijh@pmo.ac.cn\\

\noindent 0275-1062/01/\$-see front matter $\copyright$ 2014 Elsevier
Science B. V. All rights reserved. 

\noindent PII: }

\enauthor{YU Liang-liang$^{1,2,3}$$^{\bigtriangleup}$\hs\hs JI
Jiang-hui$^{1,2}$$^{\bigtriangleup}$$^{\bigtriangleup}$\hs\hs WANG Su$^{1,2}$}
{\up{1}Purple Mountain Observatory, Chinese Academy of Sciences, Nanjing 210008\\
\up{2}Key Laboratory of Planetary Sciences, Chinese Academy of Sciences, Nanjing 210008\\
\up{3}University of Chinese Academy of Sciences, Beijing 100049}

\abstract{In order to obtain the substantial information
 about the surface physics
and thermal property of the target asteroid (162173) 1999 JU3, which
will be visited by Hayabusa 2 in a sample return mission, with the
Advanced Thermal Physical Model (ATPM) we
estimate the possible thermal inertia distribution over its surface,
and infer the major material composition of its surface materials.
In addition,
the effective diameter and geometric albedo are derived to be
$D_{\rm eff}=1.13\pm0.03\rm~km$, $p_{\rm v}=0.042\pm0.003$,
respectively, and the average thermal inertia is estimated to
be about $(300\pm50)\rm~J\cdot m^{-2}\cdot s^{-0.5}\cdot K^{-1}$.
According to the derived thermal inertia distribution, we infer that
the major area on the surface of the target asteroid may be covered by
loose materials, such as rock debris, sands, and so on, but few bare rocks
may exist in a very small region. In this sense, the sample return mission
of Hayabusa 2 is feasible, when it is performed successfully,
it will certainly bring significant scientific information
to the research of asteroids.}

\keywords{asteroids: individual: (162173) 1999 JU3, methods:
numerical}

\maketitle

\section{INTRODUCTION} 

The near Earth asteroid (NEA) (162173) 1999 JU3 (hereafter 1999
JU3 for brevity) is the target of the sample return mission of
Hayabusa 2  carried out by JAXA. According to the plan,
Hayabusa 2 will be launched in 2014, arrive at 1999 JU3 in 2018, and
return back to the Earth in 2020 with the sample obtained from the
asteroid$^{[1]}$. Early in 2003, the first Hayabusa probe of
JAXA had completed a sample return mission from the S-type NEA
Itokawa, and obtained a lot of important information about the
km-sized asteroid$^{[2]}$. The S-type asteroids with a great
quantity reside mainly in the inner part of the main asteroid belt
(MAB), while the C-type asteroids are relatively plentiful in the
middle and outer parts of the MAB. Hayabusa 2 selects 1999 JU3 as the
target, because it is a near Earth C-type asteroid that is usually
regarded as the parent body of carbonaceous chondrite,
containing much more organic materials than other types of meteorites.
Therefore, the investigation on small celestial objects of this type
may bring about substantial information about the
primordial asteroids and the early evolutionary stage of the solar system,
and provide important clues for the study of the origin and evolution of
the solar system. The probe of 1999 JU3 has a very significant
scientific value.

To successfully achieve this scientific objective, it is necessary to
obtain beforehand some fundamental knowledge about this object such
as the basic physical and thermal properties, with
which the space mission can be designed properly. As a target of
sample return mission, the thermal inertia of 1999 JU3 is of
particular interest. According to the definition of thermal inertia,
$\Gamma=\sqrt{\rho c\kappa}$ (where $\rho, c$ and $\kappa$ is the
mass density, specific heat capacity and thermal conductivity,
respectively), its value closely relates with the existence and
depth of surface soil, and therefore the feasibility of the sample
return mission. The physical properties and thermal inertia of
1999 JU3 have been investigated by Hasegawa et al.$^{[3]}$, Campins
et al.$^{[4]}$ and M\"{u}ller et al.$^{[5]}$, below we
briefly review their results.

Adopting the simplified spherical or ellipsoidal shape model and
the simplified assumption about the orientation of rotation axis,
and using the Thermal
Physical Model$^{[6-9]}$ (TPM hereafter), Hasegawa et al.$^{[3]}$
fitted in 2008 the observational data of 1993 JU3 from Akari and
Subaru, and they concluded: the effective diameter of this asteroid
is $D_{\rm eff}=0.92\pm0.12\rm~km$, the geometrical albedo is $p_{\rm
v}=0.063^{+0.020}_{-0.015}$, and the average thermal inertia is
probably larger than $500\rm~J\cdot m^{-2}\cdot s^{-0.5}\cdot
K^{-1}$.

In 2009, Campins et al.$^{[4]}$ applied the TPM to fitting the mid-infrared
data of Spitzer obtained on May 2, 2008 and studied the average
thermal inertia. In their fitting, a spherical shape was assumed,
but for the rotation axis orientation, two extreme cases were
considered. In one of them, the rotation axis is perpendicular to
the equator and it has a retrograde rotation, while the other
one is the direct rotation proposed by Abe et al.$^{[10]}$
with the rotation axis orientation to be $\lambda=331.0^\circ,
\beta=+20^\circ$ ($\lambda$
and $\beta$ are ecliptic longitude and latitude respectively). Their
conclusions can be summarized as follows: for the former rotation
model, the thermal inertia of 1993 JU3 has a small value of about
$150\rm~J\cdot m^{-2}\cdot s^{-0.5}\cdot K^{-1}$; but for the latter
case, the best-fit result is $\Gamma = (700\pm200)\rm~J\cdot
m^{-2}\cdot s^{-0.5}\cdot K^{-1}$. Besides, using the Near Earth
Asteroid Thermal Model$^{[11]}$ (NEATM), Campins et al.$^{[4]}$ also
made fitting of the mid-infrared data from Akari and
Subaru$^{[3]}$, and their best-fit results were respectively $D_{\rm
eff}=0.80\pm0.12\rm~km$, $p_{\rm v}=0.08\pm0.03$, $\eta=1.0\pm0.4$
and $D_{\rm eff}=1.13\pm0.17\rm~km$, $p_{\rm v}=0.04\pm0.02$,
$\eta=2.1\pm0.6$, where $\eta$ is the ``beaming parameter''.

Later in 2011, using the lightcurve inversion method$^{[12]}$,
M\"{u}ler et al.$^{[5]}$ obtained 84 kinds of possible rotational
 orientation and shape models of 1999 JU3. Then, considering
all the possible thermal inertia in the range $0\sim2500$ $\rm J\cdot m^{-2}\cdot
s^{-0.5}\cdot K^{-1}$, and adopting respectively these 84 models, they used the TPM to
fit the observational data from Spitzer, Akari and Subaru, and finally obtained the
best-fit rotational orientation: $\lambda=73.0^\circ$, $\beta=-62^\circ$,
rotation period: $P=7.63\pm0.01\rm~h$, as well as the best-fit effective diameter,
geometrical albedo and thermal inertia: $D_{\rm eff}=0.87\pm0.03\rm~km$, $p_{\rm v}=0.070\pm0.006$,
$\Gamma=200\sim600\rm~J\cdot m^{-2}\cdot s^{-0.5}\cdot K^{-1}$.

Although M\"{u}ller et al.$^{[5]}$ studied carefully the shape of
1999 JU3 and obtained in their work relatively precise spin
orientation, their investigation on thermal inertia gave only the
range of possible average thermal inertia. But in fact the thermal
inertia of an asteroid should have a specific distribution on its
surface, not be homogeneous everywhere. If the thermal inertia
distribution of the target asteroid can be found by data fitting, it
will offer a criterion for the selection of the sampling location
of the probe, and consequently improve the success possibility of
the space mission. Therefore, compared with studying only the
possible average thermal inertia, a study on the surface distribution of
thermal inertia of 1999 JU3 has much more scientific significance.

In this paper, we will study the distribution of thermal inertia on
the surface of 1993 JU3 and further analyze its surface properties.
First, in Section 2 we will use the 60 lightcurves (from Minor
Planet Center, MPC) to reconstruct the 3-dimensional shape of this
asteroid with the lightcurve inversion method. After the
introductions to the ATPM$^{[13-14]}$ and the fitting
process$^{[15]}$ of the mid-infrared data in Section 3, the
fitted results will be presented in Section 4. And finally, in
Section 5 and Section 6, we discuss and conclude our calculations.

\section{SHAPE MODEL}

To construct the shape model using the lightcurve inversion
method$^{[12]}$, we need at least 10 lightcurves and an initial
rotation axis orientation. We acquire 60 lightcurves from the MPC, and adopt
the rotation axis orientation proposed by M\"{u}ller et al.$^{[5]}$
($\lambda=73.0^\circ$, $\beta=-62^\circ$) that was obtained by a
thorough analysis in their work, and in this
paper we fix the rotational orientation on this value. Finally, we obtain the
best-fit 3-dimensional shape that is composed of 2016
triangular surface elements with different sizes and of 1010
vertices, as shown in Fig.~1.

It is worth emphasizing that the coordinates in Fig.~1 are only the
relative values without any specified units, i.e. Fig.~1 only
displays the relative size of 1993 JU3 (the real size will be given
later in this paper). Hereinafter, we will adopt this shape model
and apply a modified ATPM to fitting the observational data in Table~1.
\bc{\small\bf

Table 1~~ Subaru mid-infrared observations of asteroid 1999 JU3 on
2007 August 28$^{[3]}$
}\\[2mm]

\fns \tabcolsep 2mm
\begin{tabular}{cccccc}
\hline
No. & Observation time (UT) & Filter band & $\lambda$/$\rm\mu m$ & Flux/mJy & Error/mJy \\
\hline
 1  & 11:13:55 & N11.7 & 11.7 & 95.62 & 10.24 \\
 2  & 11:27:26 & N12.4 & 12.4 & 105.12& 15.33 \\
 3  & 11:36:29 & N10.5 & 10.5 & 70.07 & 7.19  \\
 4  & 11:45:34 & N9.7  & 9.7  & 75.47 & 17.20 \\
 5  & 11:52:48 & N11.7 & 11.7 & 98.72 & 8.58  \\
 6  & 12:04:29 & N8.8  & 8.8  & 61.50 & 7.03  \\
 7  & 12:15:50 & N11.7 & 11.7 & 103.41& 9.81  \\
 8  & 12:48:58 & N11.7 & 11.7 & 110.07& 13.48 \\
 9  & 13:00:29 & N12.4 & 12.4 & 100.98& 20.43 \\
 10 & 13:11:33 & N8.8  & 8.8  & 47.90 & 10.19 \\
 11 & 13:22:48 & N11.7 & 11.7 & 106.23& 11.94 \\
 12 & 13:48:00 & N11.7 & 11.7 & 84.56 & 10.58 \\
 13 & 14:00:03 & N12.4 & 12.4 & 99.62 & 21.47 \\
 14 & 14:23:00 & N8.8  & 8.8  & 52.24 & 10.21 \\
 15 & 14:35:31 & N11.7 & 11.7 & 114.79& 21.85 \\
 \hline
\end{tabular}
\ec
\vspace*{1.7mm}
\begin{figure}
\vspace*{-4mm} \center 
{
\includegraphics[scale=1.0]{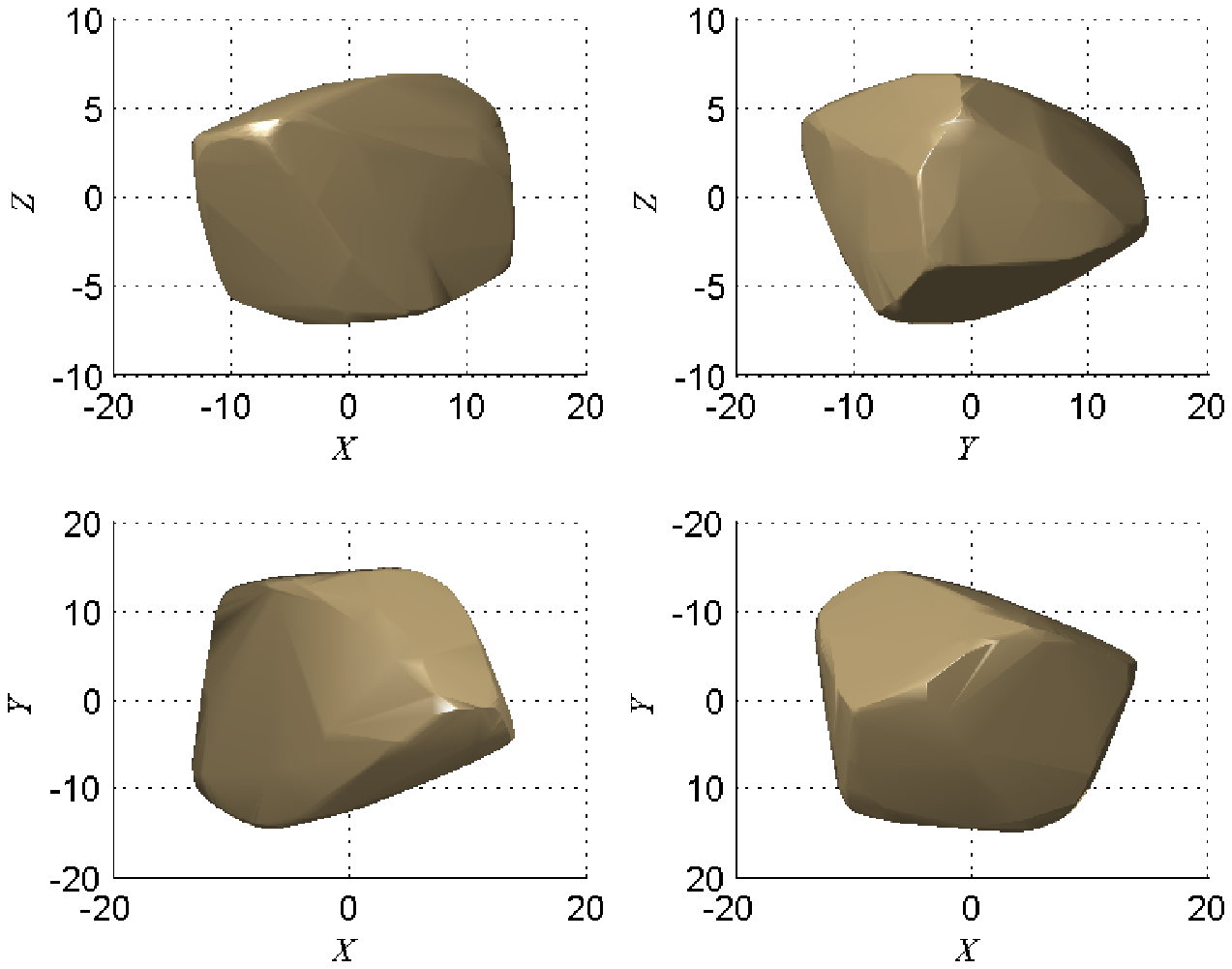}} 
{\center\footnotesize
   Fig.\,1~~ The 3D shape model of asteroid 1999 JU3
\label{shape}}

\end{figure}

\section{THERMAL MODEL AND FITTING METHOD}

\subsection{Thermal Model}

With a thermal model, we can simulate the surface
temperature distribution of the asteroid, and then deduce the
infrared radiation flux. Finally, combining these results with
observational data we can study the surface property of the
asteroid. This is the basic idea we will follow in this paper. As
the latest thermal physics model of asteroids, ATPM can be applied to
studying the thermal infrared beaming effect$^{[13-14]}$ of the rough
surface of an asteroid. However, in this paper we will focus on the
possible thermal inertia distribution on the asteroid surface, but
ignore the effect of the rough surface in the fitting process.
Therefore, we will apply directly the shape model obtained from the
lightcurve fitting, but ignore the possible pits due to the
surface roughness.

In the ATPM, an asteroid is assumed to be a polyhedron consisted of
$N$ surface elements. On each surface element exists the thermal
balance of the incident sunlight flux $q_{\rm
s}$, infrared thermal emission flux, multi-scattered sunlight flux
from other surface elements $q_{\rm scat}$, radiation heating flux from
other surface elements $q_{\rm red}$ and the 1-dimensional heat
conduction along the radial direction, which can be described by an
equation as
\begin{equation}
(1-A_{\rm B})(\mu_{0}\cdot q_{\rm s}+q_{\rm scat})+(1-A_{\rm
th})q_{\rm red}=
   \varepsilon\sigma T^{4}|_{h=0}+ \Big(-\kappa\frac{{\rm d}T}{{\rm d}h}\Big)\Big|_{h=0}\,,
\label{atpm1}
\end{equation}
where $A_{\rm B}$ and $A_{\rm th}$ are respectively Bond albedo and infrared
albedo, $\mu_{0}$ denotes the cosine of altitude of the Sun,
$\varepsilon$, $\sigma$, $\kappa$ are thermal emissivity, Stefan's
constant and heat conductivity, respectively, $T$ is temperature and
$h$ is the depth. By approximation, $q_{\rm s}$ reads$^{[16]}$:
\begin{equation}
q_{\rm s}=\frac{F_{\odot}}{d_{\rm s}^{2}}\,,
\end{equation}
with $F_{\odot}$ being the solar constant and $d_{\rm s}$ being the
distance from the asteroid to the Sun in astronomical unit (au). For
an asteroid covered with surface soil, the temperature under the soil is
assumed to be constant, i.e. exists an inner boundary condition as follows.

\begin{equation}
\frac{\partial T}{\partial h}\Big|_{h\rightarrow\infty}=0\,.
\label{atpm2}
\end{equation}
While in the surface soil, the temperature is determined by the
1-dimensional heat conduction equation:
\begin{equation}
\rho c\frac{\partial T}{\partial
t}=\kappa\frac{\partial^{2}T}{\partial h^{2}}\,. \label{atpm3}
\end{equation}
Taking Eq.(\ref{atpm1}) and Eq.(\ref{atpm2}) as the boundary
conditions, the temperature distribution on the asteroid surface and
subsurfaces can be calculated by numerically solving the
partial differential equation of Eq.(\ref{atpm3}). Suppose the heat
radiation from each surface element can be approximated by a grey
body$^{[17]}$ of emissivity $\varepsilon$, the infrared radiation
flux $F(\lambda)$ from the asteroid toward the Earth can be
calculated, provided the surface temperature distribution is known.
\begin{equation}
F(\lambda)=\sum^{N}_{i=1}\varepsilon f_{i}B(\lambda,T_{i})\,,
\end{equation}
where $B(\lambda,T_{i})$ is Planck monochromatic radiation flux, and
$f_{i}$ stands for the view factor of each surface element with
respect to the Earth$^{[18]}$. The view factor is defined as
\begin{equation}
f_{i}=A_{i}\frac{\emph{\textbf{n}}_{i}\cdot{\bf\Delta}}{\pi d_{\rm
e}^2}\,,
\end{equation}
where $A_{i}$ and $\emph{\textbf{n}}_{i}$ are the area, unit normal
vector of the $i$th surface element, while $\bf\Delta$ and $d_{\rm
e}$ are the unit vector pointing to and the distance to the Earth
from this surface element, respectively.

\subsection{Fitting Method}

As mentioned above, the shape model of 1999 JU3 has been obtained
from the lightcurve inversion, although without knowing its size. We
will derive its surface thermal inertia distribution through fitting
the mid-infrared data using the thermal physics model. Since our
shape model is a polyhedron consisting of $N=2016$ surface elements,
the thermal inertia distribution on the surface can be described
naturally by $\Gamma(i)$ with $i=1,2,\cdots N$ indicating each of
the surface elements. Hence, the free parameters in the fitting
process now include the effective diameter, albedo, and $\Gamma(i)$.

In fact, between the effective diameter $D_{\rm eff}$ and the
geometrical albedo $p_{\rm v}$$^{[19]}$ an empirical relation
exists:
\begin{equation}
D_{\rm eff}=\frac{1329\times 10^{-H_{\rm v}/5}}{\sqrt{p_{\rm
v}}}~(\rm km)\,, \label{Deff}
\end{equation}
where $H_{\rm v}$ is the absolute magnitude. Moreover, the
geometrical albedo $p_{\rm v}$ and the Bond albedo $A_{\rm B}$ are
related by the following equation$^{[20]}$:
\begin{equation}
p_{\rm v}=\frac{A_{\rm B}}{q}\,, \label{pvA}
\end{equation}
where $q$ is the phase integral$^{[21]}$. Both the effective diameter
and Bond albedo are necessary quantities in the fitting process. They are
related to each other through the geometrical albedo, thus either of
them can be regarded as the free parameter.
Consequently, in the fitting process we have only two free parameters, i.e. the effective
diameter (or equivalently the geometrical albedo) and $\Gamma(i)$.

Because we focus on the possible thermal inertia distribution
$\Gamma(i)$, but not the average thermal inertia $\Gamma$, and this
thermal inertia distribution should not be arbitrarily valuated, a possible
algorithm to evaluate the thermal inertia $\Gamma(i)$ shall be
searched in this paper. Due to the fact that the shape is obtained
from the lightcurve inversion, the different sizes of surface elements
reflect in a sense the different reflectivity in the different region on the
asteroid surface. A larger surface element indicates that there are
more direct reflections (mirror-like reflections) in the corresponding
area, namely, in this area the surface is more smooth and the
grains of surface material are relatively small. Contrarily, a
small surface element corresponds to a relatively rough region,
possibly decorated with protruded rocks. Hence, the surface
distribution of thermal inertia $\Gamma(i)$ could be related
approximately to the area of the corresponding surface element
$A_{i}$.

In 2007, Delbo et al.$^{[22]}$ obtained from their statistics an
approximate relation between the effective diameter $D_{\rm eff}$
and the average thermal inertia $\Gamma$ of an asteroid:
\begin{equation}
\Gamma=d_{0}\big(\frac{D_{\rm eff}}{1~ \rm km}\big)^{-\xi}\,,
\label{gmaD}
\end{equation}
where $d_{0}=300\pm47\rm~J\cdot m^{-2}\cdot s^{-0.5}\cdot K^{-1}$,
$\xi=0.48\pm0.04$. This relation reveals the fact that a larger
asteroid has a smaller surface thermal inertia. Inspired by this, we
suppose that there could be a similar relation between $\Gamma(i)$
and $A_{i}$ as below.
\begin{equation}
\Gamma(i)=\alpha A_{i}^{-\beta}\,. \label{gmaA}
\end{equation}
Of course, there is no necessary connection between the parameters
$\alpha$, $\beta$ in Eq.(\ref{gmaA}) and the parameters $d_{0}$, $\xi$ in
Eq.(\ref{gmaD}). The best-fit values of the parameters $\alpha$ and $\beta$ can be
determined by fitting the observational data. Because $A_{i}$
comes from directly the shape model, it indicates only the relative area,
the value of $\beta$ depends on the scale adopted in the shape model.

For the thermal inertia distribution given by Eq.(\ref{gmaA}),
$\Gamma(i)=\alpha=\rm const$ if $\beta=0$, corresponding exactly to
the situation of average thermal inertia, i.e. all surface elements
have the same thermal inertia. But the thermal inertia for the surface
elements with a large area should be relatively small, hence $\beta \geq
0$. Especially, we find in our investigation that the thermal inertia
distribution $\Gamma(i)$ displays an evidently non-normal distribution
if $\beta$ is larger than a specific value. It doesn't make sense in
physics. Therefore, an upper limit should be set for $\beta$. For
the shape model in Fig.~1, we confine the value of $\beta$ in the
range of $0\sim 0.1$. As $\alpha$ is the average thermal inertia, we
take $\alpha=200\sim600\rm~J\cdot m^{-2}\cdot s^{-0.5}\cdot K^{-1}$,
which is the average thermal inertia of 1999 JU3 given by M\"{u}ller
et al.$^{[5]}$. To simplify the fitting process, we choose only
three values of $\beta=0.00, 0.05, 0.10$ and $\alpha = 200 \sim 600
\rm~J \cdot m^{-2} \cdot s^{-0.5} \cdot K^{-1}$, totally 24 thermal
inertia distributions, to discuss in this paper.

To derive the best-fit $\Gamma(i)$, we need to determine the
best-fit values of $D_{\rm eff}$ and $p_{\rm v}$ at the same time.
For each distribution $\Gamma(i)$, starting from an initial Bond
albedo $A_{\rm B, initial}=0.01$, we first calculate the theoretical
radiation flux using the ATPM program, then select a series values of
effective diameter $D_{\rm eff}$ (or equivalently geometrical albedo
$p_{\rm v}$), and introduce the correction factor FCF$^{[23]}$, namely
\begin{equation}
{\rm FCF}=\frac{1-A_{\rm B,now}}{1-A_{\rm B,initial}}\,, \label{FCF}
\end{equation}
to fit the observational data, and finally find the best-fit effective
diameter $D_{\rm eff}$ and geometrical
albedo $p_{\rm v}$. The best-fit value is defined to let the
difference between the theoretical flux and the observational
flux$^{[23]}$, namely

\begin{equation}
L^2=\sum_{i=1}^{n} \Big[ \frac{F(\lambda_i,{\rm
model})-F(\lambda_i,{\rm obs})}{\sigma_{i}} \Big]^{2} \label{FCF},
\end{equation}
be minimum. In this equation, the subscript $i$ indicates
an observation, $\lambda_{i}$ and $\sigma_{i}$ stand for the
wavelength and error of this observation. Apparently, $D_{\rm eff}$
and $p_{\rm v}$ are the values correspond to the best-fit
$\Gamma(i)$.

\begin{figure}
\vspace*{4mm}
\center
{\includegraphics[scale=0.7]{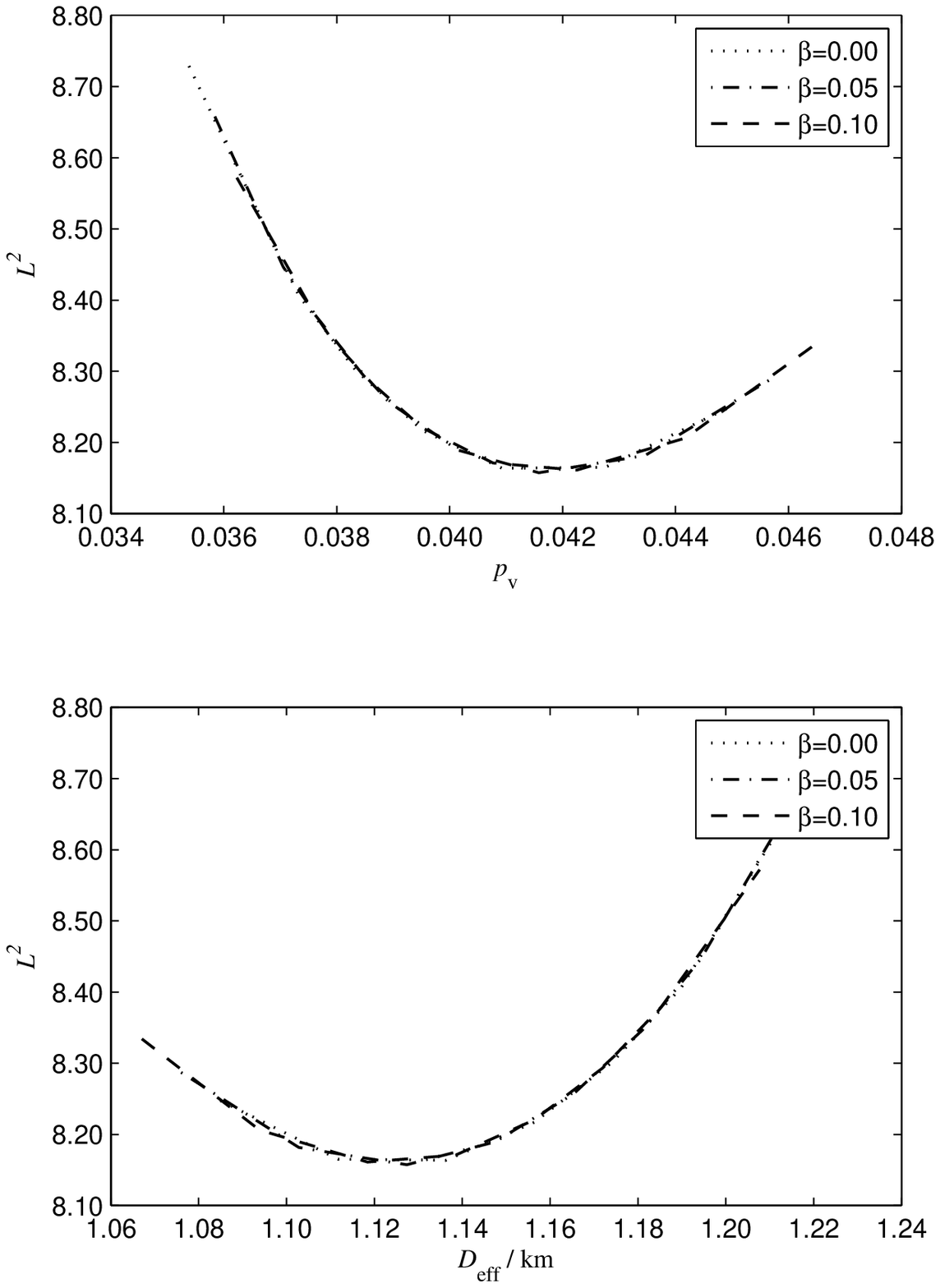}}

{\center\footnotesize
   Fig.\,2~~ The fitting of thermal inertia of 1999 JU3:  $L^{2}-p_{\rm v}$ and $ L^{2}-D_{\rm eff}$ curves
}
\end{figure}

\section{RESULTS}

\subsection{Effective Diameter and Geometrical Albedo}
In our simulations, we have tried to fit the Subaru infrared data
using 24 different kinds of thermal inertia distribution $\Gamma(i)$, and
derived the respective effective diameter $D_{\rm eff}$, geometrical albedo
$p_{\rm v}$ and the corresponding $L^{2}$ value. To dig out the best-fit
result, i.e. the situation corresponding to the smallest $L^2$, from the 24
situations, we plot the curves of $L^{2}- D_{\rm eff}$ and $L^{2} -
p_{\rm v}$ as in Fig.~2. The lowest points are the expected $D_{\rm
eff}$ and $p_{\rm v}$.

In Fig.~2 the minima at $p_{\rm v}=0.042$, $D_{\rm
eff}=1.13~{\rm km}$ are the best-fit results. Considering the
unavoidable errors, we take 1\% of the minimum $L^{2}$ as the error
and obtain the ranges of best-fit geometrical albedo and effective
diameter: $p_{\rm v}=0.042\pm0.003$, $D_{\rm
eff}=1.13\pm0.03\rm~km$.

It is worth noting that the best-fit $D_{\rm eff}$ and $p_{\rm v}$
obtained here depend only on the value of $\alpha$, but nearly not
on $\beta$, implying that in the calculations of $D_{\rm eff}$ and
$p_{\rm v}$, we need only the average thermal inertia in the thermal
model and no other results will appear even if the thermal inertia
distribution is taken into account.

\subsection{Thermal Inertia Distribution and Surface Characteristics}

\no{4.2.1~~Thermal inertia distribution}

To find the best-fit thermal inertia distribution $\Gamma(i)$, we locate
the minimum of the $L^{2}- \Gamma(i)$ curve (as shown in Fig.~3). The dotted
line in Fig.~3 ($\beta=0.00$), corresponding to just the degenerated
case of average thermal inertia $\Gamma(i)=\alpha$, has its minimum
at $\alpha\approx300\rm~J\cdot m^{-2}\cdot s^{-0.5}\cdot K^{-1}$.
Other two curves for $\beta=0.05$ and $\beta= 0.1$, corresponding to
the distribution described in Eq.(\ref{gmaA}), attain their minima
also at $\alpha\approx300\rm~J\cdot m^{-2}\cdot s^{-0.5}\cdot
K^{-1}$, implying that $\alpha$ is a constant, i.e. the average
thermal inertia of 1999 JU3 is about $300\rm~J\cdot m^{-2}\cdot
s^{-0.5}\cdot K^{-1}$. Taking account of the 1\% error of the
minimum of $L^2$, the average thermal inertia is
$(300\pm50)\rm~J\cdot m^{-2}\cdot s^{-0.5}\cdot K^{-1}$.
\begin{figure}

\center {\includegraphics[scale=0.8]{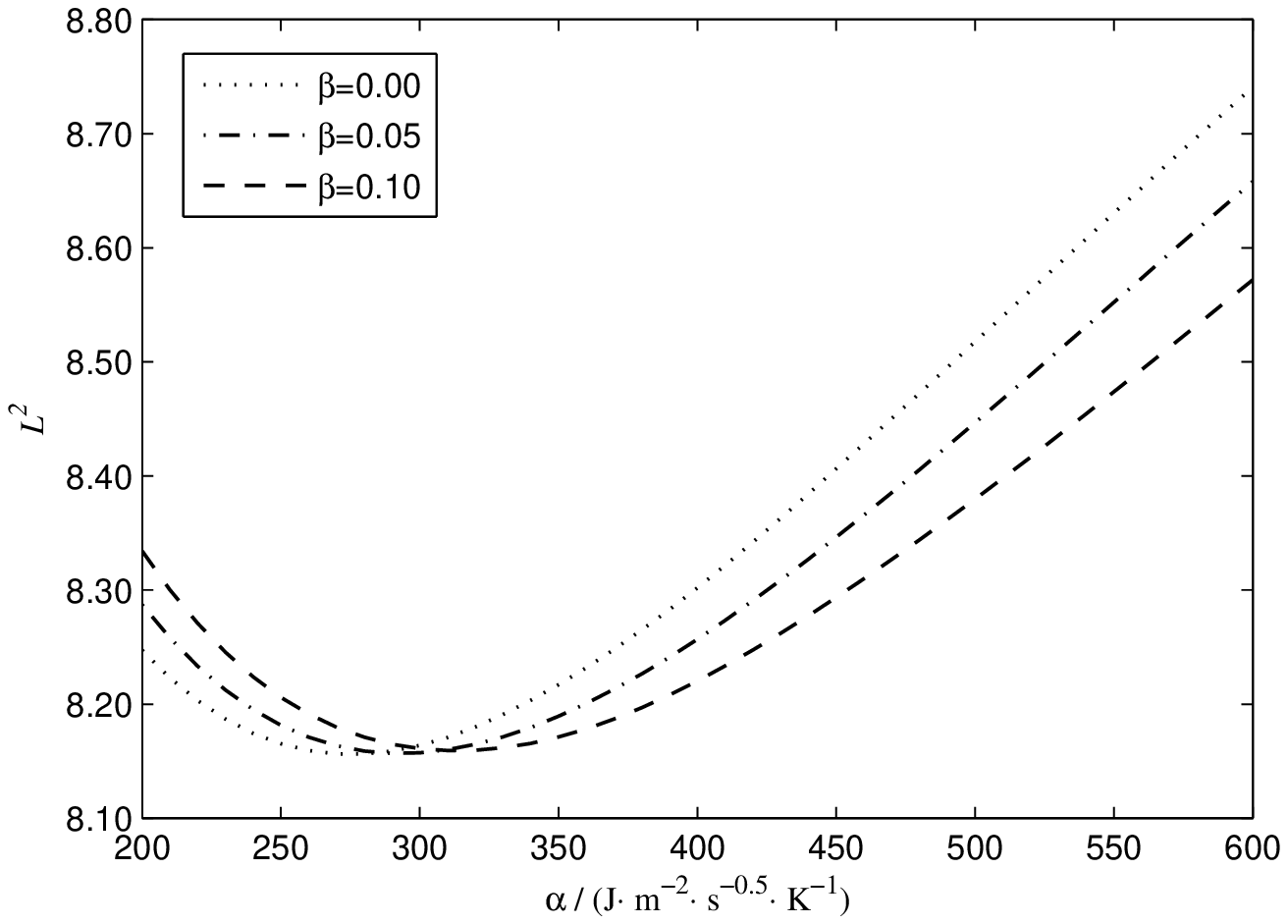}}

{\center\footnotesize
   Fig.\,3~~ The fitting results of 1999 JU3: $L^{2}-\alpha$ curve
}

\end{figure}

For these 3 best-fit thermal inertia distributions, the goodness of
fitting to the Subaru data are checked by plotting the infrared
spectra in Fig.~4 and comparing them with the observation. The
consistence between the observation and the fitted spectra
indicates evidently that all three thermal inertia distributions of
$\beta=0.00, 0.05, 0.10$ may fit very well the Subaru infrared spectrum.
Additionally, we
also plot the infrared rotation curve of 1999 JU3 at $\lambda=$11.7\,$\mu$m
for all these three cases in Fig.~5, indicating again that these thermal inertia
distributions can fit the infrared rotation curve quite well.
Therefore, we conclude that any thermal inertia distributions
satisfying Eq.(\ref{gmaA}) with $\alpha=300\rm~J\cdot m^{-2}\cdot
s^{-0.5}\cdot K^{-1}$, $\beta=0.0\sim 0.1$ may fit nicely the Sabaru
data of 1999 JU3.

\no{4.2.2~~ Analysis on surface characteristics}

As another limiting situation, the distribution $\Gamma(i)$ of
$\alpha=300\rm~J\cdot m^{-2}\cdot s^{-0.5}\cdot K^{-1}$,
$\beta=0.10$, as shown in Fig.~6, can be used to give an estimation
of the surface characteristics of 1999 JU3.
\newpage
\begin{figure} 

\center {\includegraphics[scale=0.8]{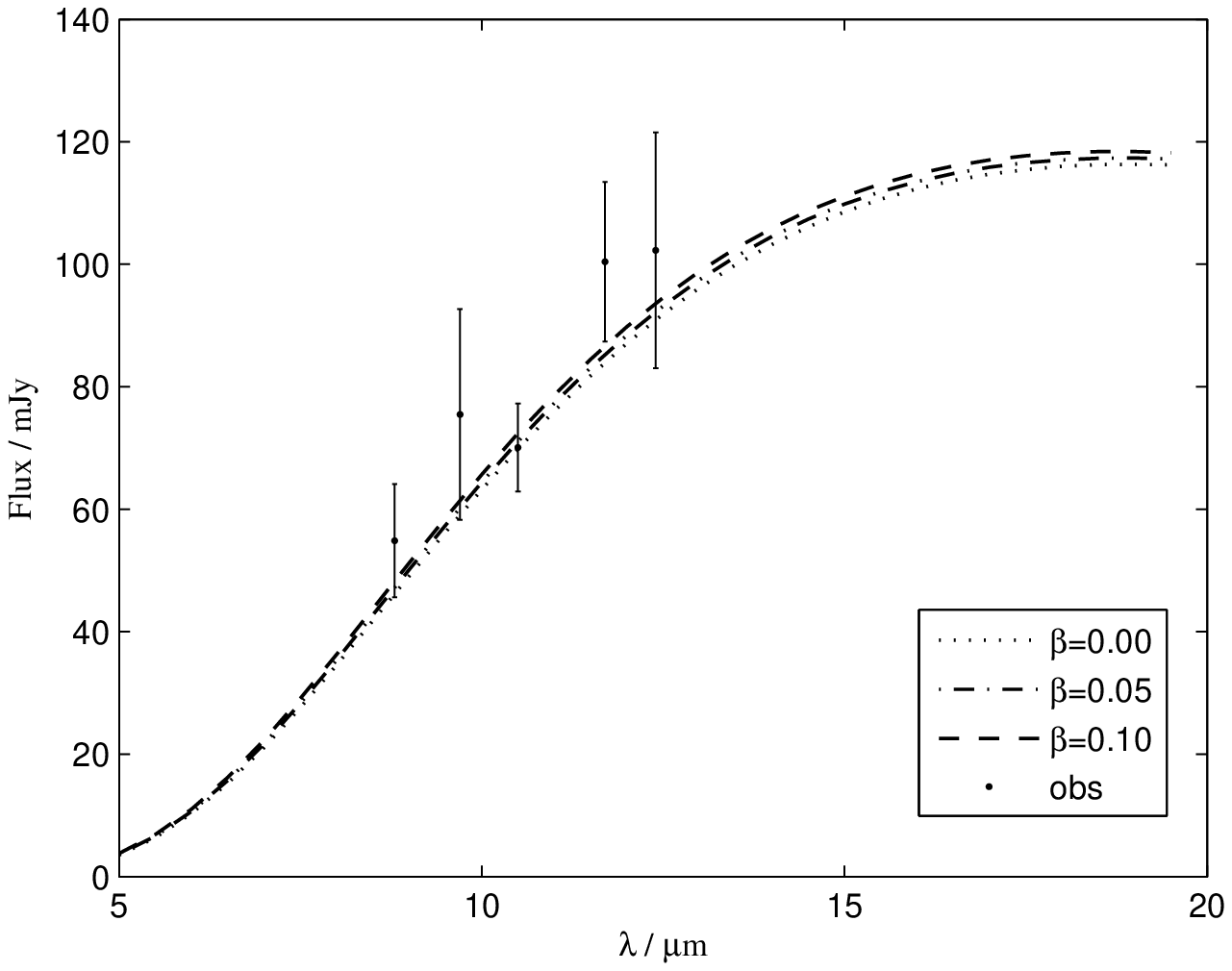}}

{\center\footnotesize
   Fig.\,4~~ The infrared spectrum of 1999 JU3 observed by Subaru on 2007 August 28 and the theoretically fitted results
}

\end{figure}

\begin{figure} 

\center {\includegraphics[scale=0.8]{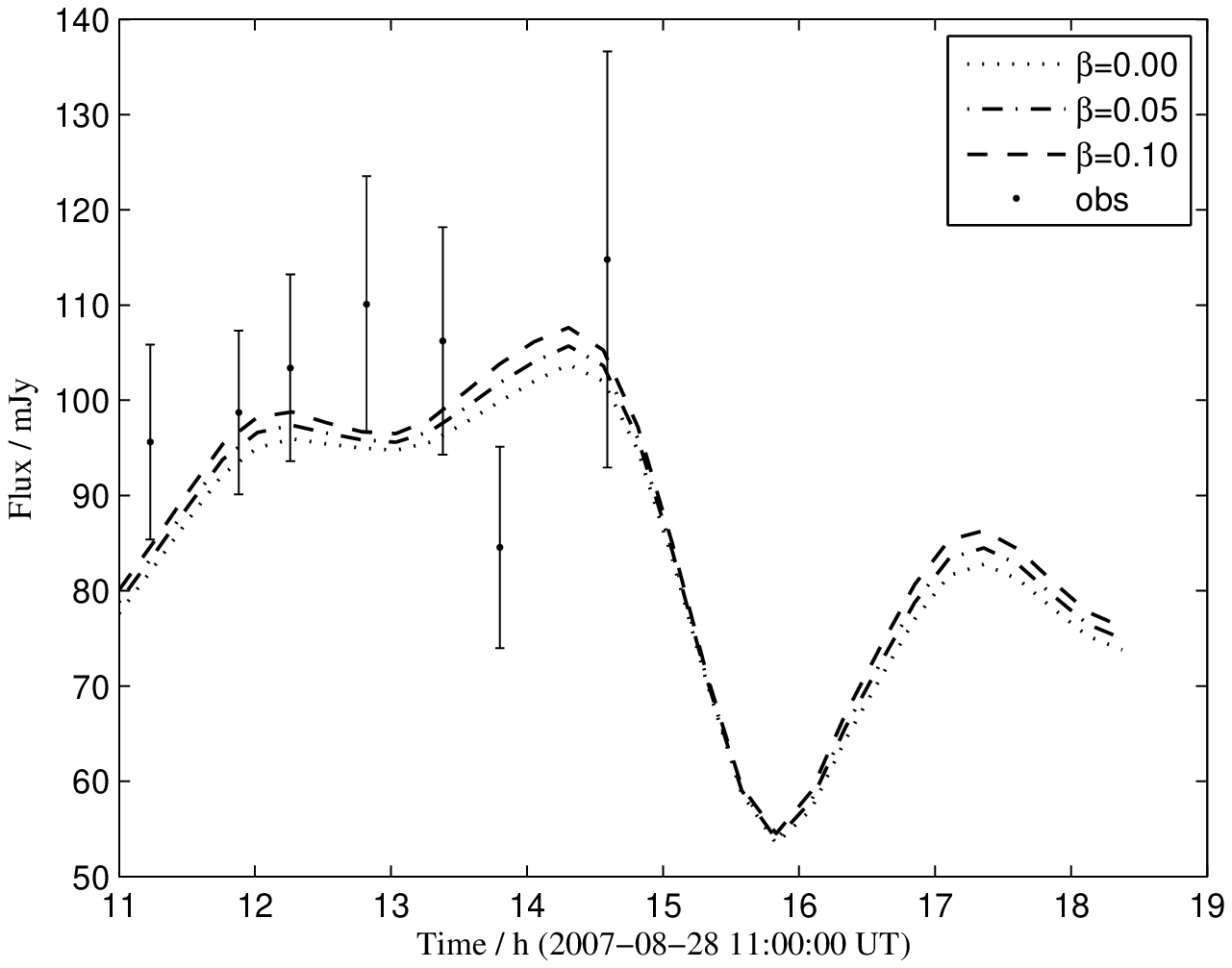}}

{\center\footnotesize
   Fig.\,5~~ The light curve of 1999 JU3 at $\lambda=11.7$ $\rm\mu m$ on 2007 August 28 and the theoretically fitted results
}

\end{figure}

According to this distribution $\Gamma(i)$, the thermal inertia in
most area on the surface of this asteroid is as small as
$200\sim400\rm~J\cdot m^{-2}\cdot s^{-0.5}\cdot K^{-1}$, in some
area it is $400\sim1000\rm~J\cdot m^{-2}\cdot s^{-0.5}\cdot K^{-1}$
and in a tiny fraction it can be as high as around $1~200\rm~J\cdot
m^{-2}\cdot s^{-0.5}\cdot K^{-1}$. The thermal inertia depends
sensitively on the surface material. Investigations show that the
thermal inertia of small dust particles is only about $30\rm~J\cdot
m^{-2}\cdot s^{-0.5}\cdot K^{-1}$, for the lunar regolith
(containing a layer of incohesive rock detritus) it is around
$50\rm~J\cdot m^{-2}\cdot s^{-0.5}\cdot K^{-1}$, for coarse gravel
it is $400\rm~J\cdot m^{-2}\cdot s^{-0.5}\cdot K^{-1}$, while for
bare rocks the thermal inertia can be as high as $2~500\rm~J\cdot
m^{-2}\cdot s^{-0.5}\cdot K^{-1}$. Thus the thermal inertia
distribution in Fig.~6 tells us that the surface of 1999 JU3 is
largely covered by loose materials, with only very small fraction of
area decorated by bare rocks.
\vspace{2mm}
\begin{figure} 

\center {\includegraphics[scale=0.82]{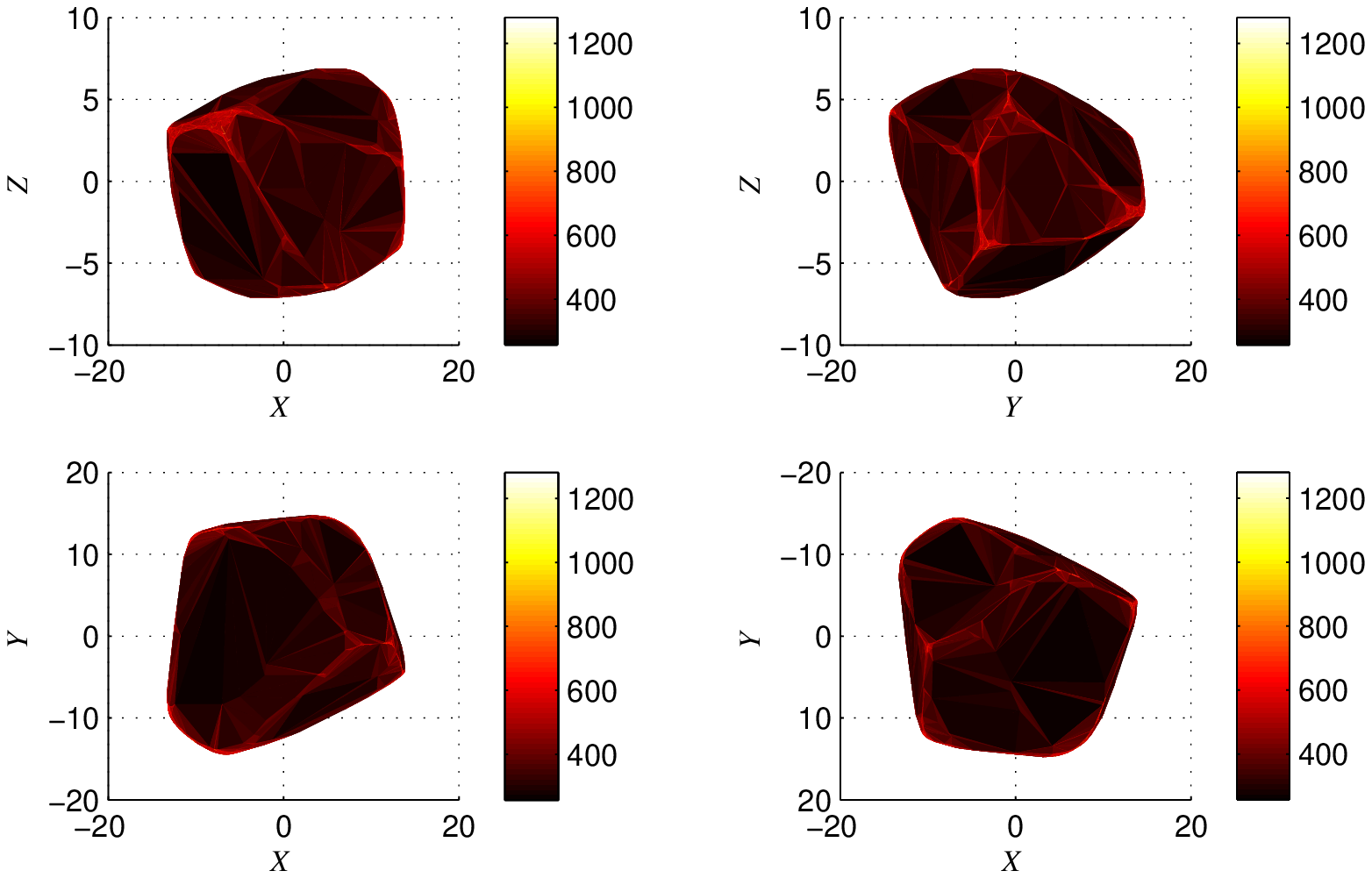}}

{\center\footnotesize
   Fig.\,6~~ The possible thermal inertia distribution of 1999 JU3. $\alpha=300\rm~J\cdot m^{-2}\cdot s^{-0.5}\cdot K^{-1}$, $\beta=0.10$.
}

\end{figure}

\section{DISCUSSIONS}

\no Combining with the mid-infrared observations, the thermal
physics model is a very effective method to study the surface physics and
thermal properties of asteroids. How advanced is the thermal
physics model and how accurate are the observational data influence
the accuracy of results. In this paper, we have applied the latest
and the most advanced model, but unfortunately the observational data
are tightly restricted. The only one set of public Subaru infrared
data puts restrictions on our results and conclusions.

Although the observational data are restricted, we have performed the
fitting on this set of observational data with the advanced thermal physics model.
Our results about the
asteroid 1999 JU3 (geometrical albedo $p_{\rm v}=0.042\pm0.003$,
effective diameter $D_{\rm eff}=1.13\pm0.03\rm~km$) are different
from what M\"{u}ller et al.$^{[5]}$ obtained, but are quite close to
the results given by Campins et al.$^{[4]}$, which were obtained by
applying the NEATM to fitting the Subaru data. This shows clearly that the
difference from M\"{u}ller et al.$^{[5]}$ arises mainly from the
restriction of observational data.

The average thermal inertia we got from our calculation is
$300\rm~J\cdot m^{-2}\cdot s^{-0.5}\cdot K^{-1}$, consistent with
the result obtained by M\"{u}ller et al. This is a little different
from the inference made by Hasegawa et al.$^{[3]}$, because we have
adopted a quite different fitting method from theirs, and also because
we have used a 3-dimensional shape model derived from lightcurve
inversion, which is closer to the real shape than the simple sphere
model adopted by Hasegawa et al. It should be noted that our results
are compatible very well with the approximate relation between the
thermal inertia and the effective diameter that is proposed by Delbo
et al.$^{[22]}$ after a careful statistical analysis. This
relation is illustrated in Fig.~7. Moreover, the spin axis orientation
is just taken from the result given by M\"{u}ller et al.$^{[5]}$,
thus the calculated average thermal inertia is consistent
with their result. Surely, other spin axis orientations may lead to
different results, since the spin of an asteroid can significantly
influence the fitting of thermal inertia. Without doubt, the more
accurate spin axis orientation, more observational data and more
sophisticated thermal physics model, will bring us a more accurate
thermal inertia estimation.
\vspace{2mm}
\begin{figure} 
\center {\includegraphics[scale=0.65]{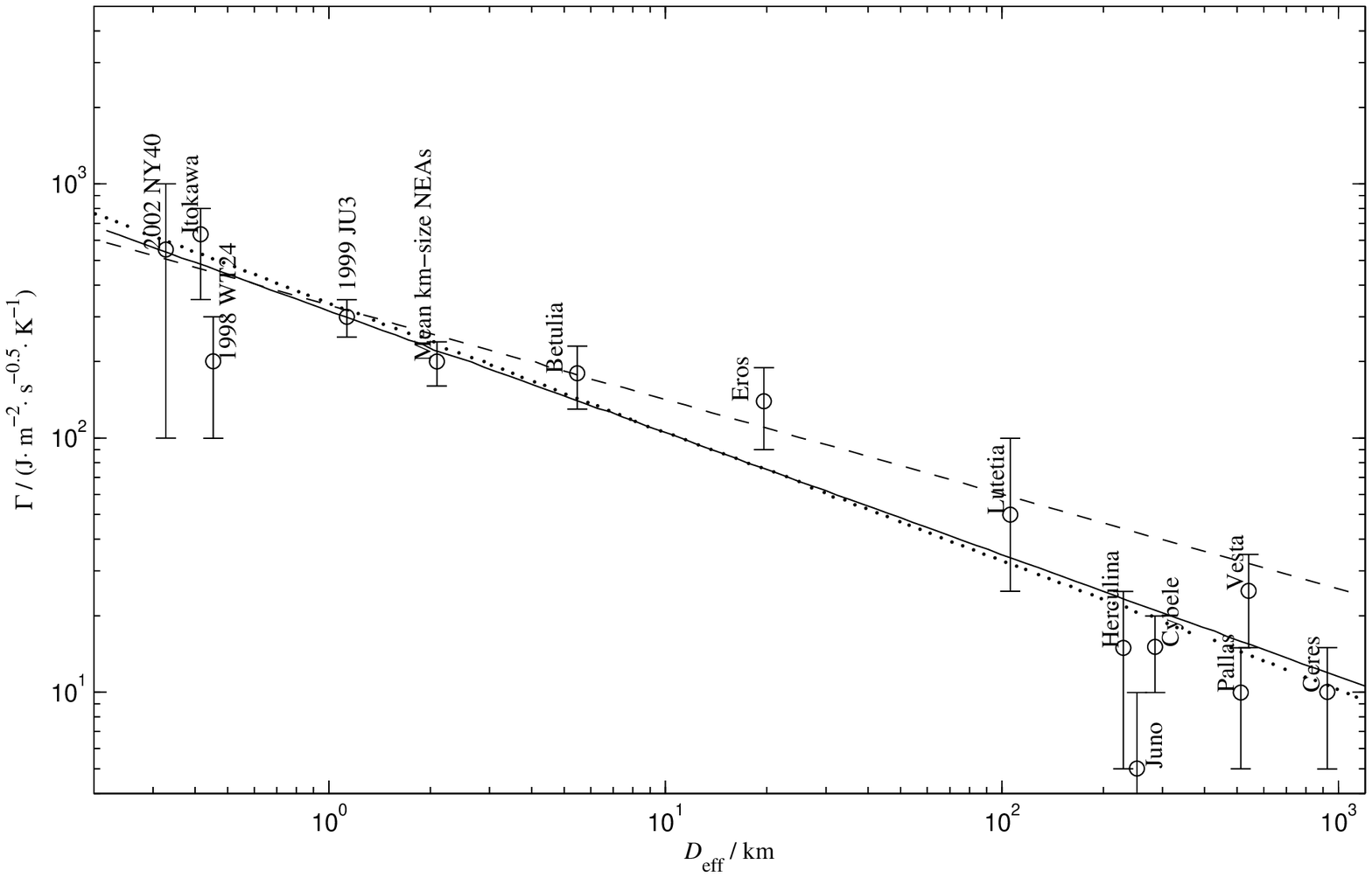}}

{\center\footnotesize
   Fig.\,7~~ The approximate relation between thermal inertia and effective diameter$^{[22]}$, where the derived values of 1999 JU3
   in this work are also listed.
}

\end{figure}

Although there are different constrains and limitations, some
related studies, e.g. estimating the range of surface thermal
inertia, are still feasible in the conditions of currently available data and
most advanced model. As the target asteroid of Hayabusa 2 sample
return mission, it is very important to obtain knowledge about the
surface characteristics of 1993 JU3. The results presented in this paper support the
technique feasibility of Hayabusa 2 mission. However, we would like
to stress that the adopted thermal inertia distribution is given by
Eq.(\ref{gmaA}), it may be not exactly the real distribution on
the surface of 1999 JU3, but it is a reliable distribution consistent
with the observational data and all the physics we have
known. If someday a more realistic thermal inertia distribution could
be obtained from more observations and some more sophisticated models,
these results would be of greater scientific value.

\section{CONCLUSIONS}

\no Based on the spin axis orientation of 1999 JU3 given by M\"{u}ller
et al., we construct the shape model using the lightcurve inversion
method. Applying the modified ATPM model, we nicely fit the Subaru
mid-infrared data. And our main results can be summarized as follows.

(1) For the asteroid 1999 JU3, its effective diameter is about
$D_{\rm eff}=1.13\pm 0.03$ km, its geometrical albedo is $p_{\rm
v}=0.042 \pm 0.003$, and the average surface thermal inertia is
$(300\pm50)\rm~J \cdot m^{-2} \cdot s^{-0.5} \cdot K^{-1}$.

(2) In most area on the surface of 1999 JU3, the thermal inertia is
$200\sim 400\rm~J \cdot m^{-2}\cdot s^{-0.5} \cdot K^{-1}$, in some
area it is $400\sim 1000\rm~J \cdot m^{-2} \cdot s^{-0.5} \cdot
K^{-1}$, and in a tiny fraction of the surface it is as high as
$1~000\rm~J \cdot m^{-2} \cdot s^{-0.5} \cdot K^{-1}$. We derive
from these values that 1999 JU3 is mainly covered by loose materials
like rock detritus and fine sands, and some area on its surface is
covered by the mixture of coarse sands and pebbles, with only very
tiny area being decorated with big bare rocks.

These results support the feasibility of the sample return mission of
Hayabusa 2 to 1999 JU3, which will bring us more scientific information
about asteroids.

\end{document}